\documentstyle[prl,twocolumn,aps,epsf]{revtex}
\def\T3{${\mathcal{T}}_3$}
\def\PRL{Phys. Rev. Lett.}
\def\PRB{Phys. Rev. B}

\begin{document}

\twocolumn[\hsize\textwidth\columnwidth\hsize\csname @twocolumnfalse\endcsname
\draft
\tolerance 500

\title{Aharonov-Bohm cages in the GaAlAs/GaAs system}
\author{C. Naud$^{1}$, G. Faini$^{1}$, D. Mailly$^{1}$, 
J. Vidal$^{2}$, B. Dou\c cot$^{3}$, G. Montambaux$^{4}$, A. Wieck$^{5}$ and
D. Reuter$^{5}$
}
\address{$^{1}$ Laboratoire de Photonique et de Nanostructures 
196, avenue Henri Rav\'era BP 29 F-92222 Bagneux cedex  France \\
$^{2}$ Groupe de Physique des Solides, CNRS UMR 7588,
Universit\'{e}s Paris 6 et 7,\\
2, place Jussieu, 75251 Paris Cedex 05 France\\
$^3$ Laboratoire de Physique Th\'{e}orique et
Hautes \'Energies, CNRS UMR 7589, Universit\'{e}s Paris 6 et 7,\\ 
4, place Jussieu, 75252 Paris Cedex 05 France\\
$^4$ Laboratoire de Physique des Solides, CNRS UMR 8502, Universit\'e
Paris Sud, B{\^a}t. 510, 91405 Orsay, France\\
$^5 $Lehrstuhl f\"ur Angewandte Festkoerperphysik Ruhr-Universit\"at Bochum\\ 
NB 03 / 58 Universit\"atsstrasse 150 D-44780 Bochum Germany
}

\maketitle

\begin{abstract}
Aharonov-Bohm oscillations have been observed in a lattice formed by a two dimensional rhombus tiling. This
observation is in good agreement with a recent theoretical calculation of the energy spectrum of this so called \T3
lattice.  We have investigated the low temperature magnetotransport of the \T3 lattice realized in the GaAlAs/GaAs
system. Using an additional electrostatic gate, we have studied the influence of the channel number on the
oscillations amplitude. Finally, the role of the disorder on the strength of the localisation is theoretically
discussed.
\end{abstract}


\vskip2pc]

\section{Introduction}

\begin{figure}
\centerline{\epsfxsize=50mm
\epsffile{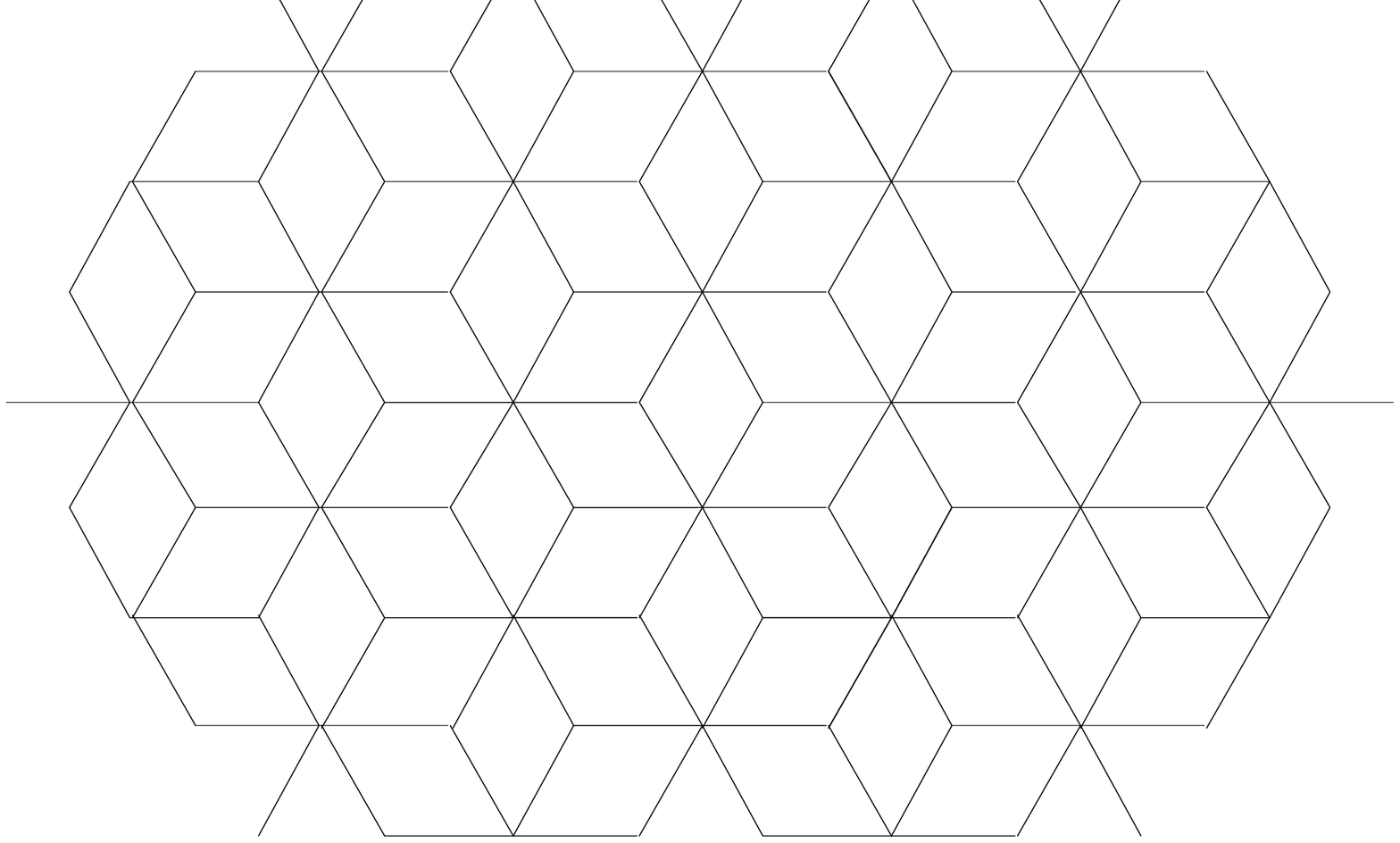}}
\vspace{2mm}
\centerline{\epsfxsize=28mm
\epsffile{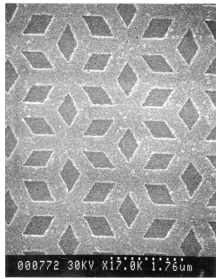}}
\vspace{2mm}
\caption{\textbf{top :} A piece of the \T3 lattice.
\textbf{bottom :} Electron microscope view of the aluminium mask used to etch 
the 2DEG. The nominal width of the wires defining the network is $0.4$ $\mu$m 
whereas their length is about $1$ $\mu$m.}
\label{lattice}
\end{figure}

\begin{figure}
\centerline{\epsfxsize=60mm
\epsffile{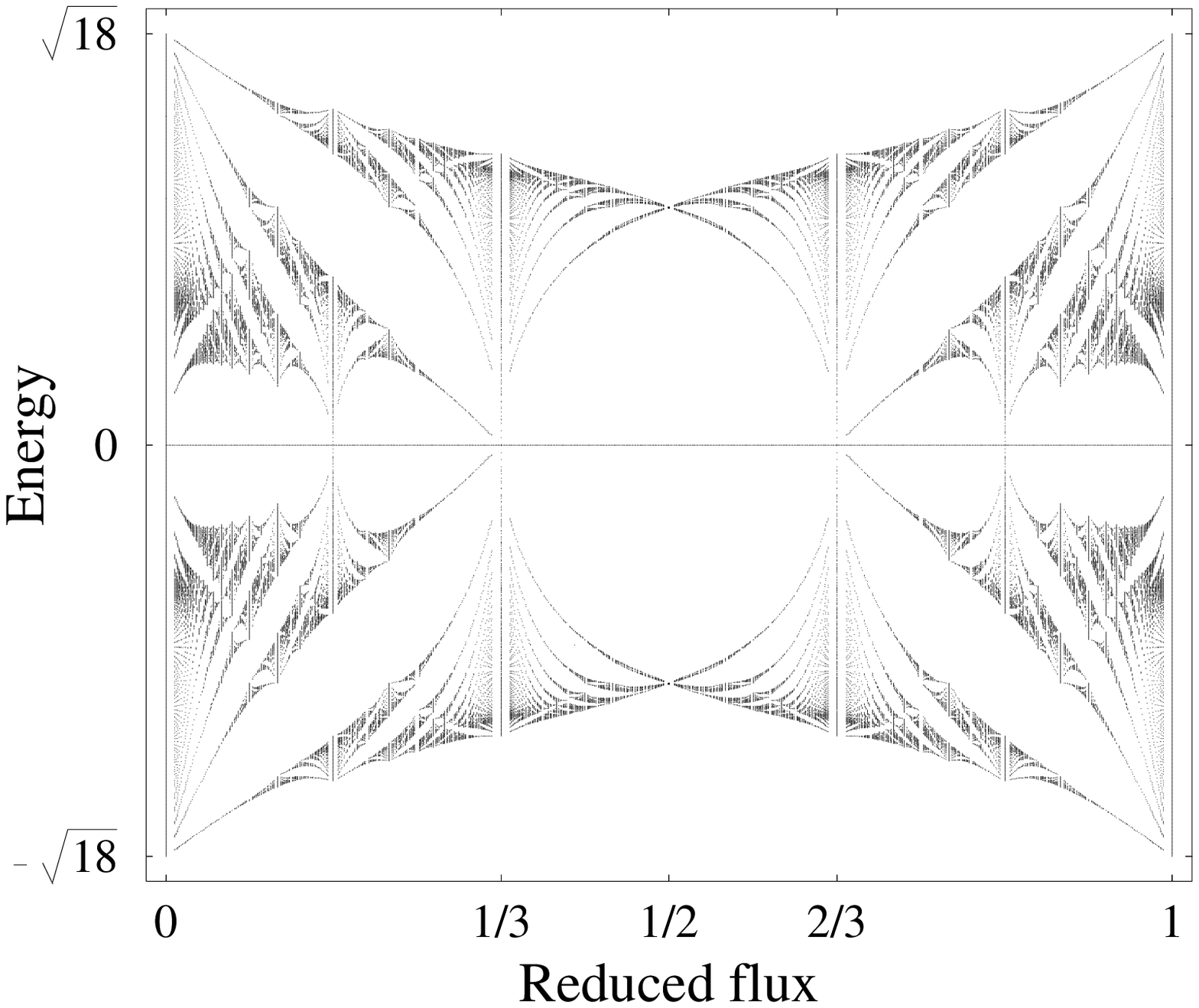}}
\caption{Energy spectrum of the \T3 lattice versus reduced flux.}
\label{spectrum}
\end{figure}

The spectral properties of an electron in a periodic lattice in the  presence of a magnetic field are of
particular interest. Indeed, the frustration between the spatial periodicity and the one induced by the magnetic
field gives rise to spectacular energy spectra versus the reduced flux $f=\phi/ \phi_{0}$, where $\phi$ is the
magnetic flux per elementary cell and $\phi_{0}$ is the flux
 quantum. A  well-known example is the Hofstadter butterfly \cite{hof}. Recently, the study of a bipartite
two-dimensional rhombus tiling,  named \T3, (shown in figure (\ref{lattice})) has attracted a lot of attention
\cite{vidal}. 
%
%
%%%%%%%%%%%%%%%%%%%%%%%%%%%%%%%%%
%%%%%%%%%%%%%%%%%%%%%%%%%%%%%%%%%
%
% 
For $f=1/2$, its energy spectrum is reduced to three  degenerate discrete levels  (see figure
(\ref{spectrum})). The whole lattice is then similar to a unique super-atom with localized electrons. This
behavior is very different from the square lattice one where for any rational value of the reduced flux the
energy spectrum is continuous. 
%
%
%%%%%%%%%%%%%%%%%%%%%%%%%%%%%%%%%
%%%%%%%%%%%%%%%%%%%%%%%%%%%%%%%%%
%
%

This localization for the \T3 lattice can be easily understood in terms of Aharonov-Bohm (AB)
effect. Indeed, in order to propagate through the lattice, an electron must travel through AB loops. For $f=1/2$ 
interferences are destructive and the electron  is then confined in a so-called AB cage. This tight-binding
calculation has received an experimental confirmation in the temperature dependence of the critical current of a
superconducting network in the \T3 geometry. In addition, the  unusual nature of the state at half flux quantum
has been revealed by a strong reduction of the critical current and by a very disordered vortex pattern through
decoration experiments \cite{pannetier,pannetier2}.

  The first experimental evidence of AB cages in a normal metal network tailored in a high mobility two
dimensionnal electron gas (2DEG) has recently been reported \cite{naud}. The low temperature magnetoresistance 
measurements show  clear $h/e$ oscillations in arrays of $2500$ cells. Experiments performed on  square lattices
of similar size do not show such a behavior . The temperature dependence of the $h/e$ peak amplitude of the
Fourier transform for both the \T3 lattice and the single rhombus have been compared. For a single loop the
amplitude of the AB signal is expected to follow a $T^{-1/2}$ law as long as the size of  the loop $L$ is smaller
than the phase coherence length $L_{\Phi}$, because of temperature averaging. When the temperature is such that
$L>L_{\Phi}$ the AB  signal falls down exponentially \cite{wasburn}. Thus, the size of a cage being larger than
that of a single loop, one expects a cutoff temperature significantly smaller for the \T3 as compared to the one
for a single rhombus. The experimental results show that this critical temperature is around $1K$ for a rhombus
whereas it is below $100$ mK for the \T3. If one assumes a $T^{-1/3}$ temperature dependence for $L_{\Phi}$ in
GaAlAs/GaAs 1D systems \cite{pepper}, the ratio of the measured critical lengths is $2.7$, in good agreement with
the geometrical dimensions yielding a ratio of $3$. More strikingly, at high magnetic field, $h/2e$ oscillations
appear whose amplitude can be much higher than the fundamental period. Such an amplitude dismisses a simple
interpretation in terms of harmonics generation. Moreover, the temperature dependence of the $h/2e$ oscillations
is similar to that of the $h/e$ one. This also indicates that the characteristic lengths associated  with the two
periodicities are alike and rules out any interpretation in terms of harmonics. One may also think of
Aronov-Altshuler-Spivak (AAS) oscillations when dealing with an $h/2e$ signal. But, due to the aspect ratio of the
sample, the AAS signal should vanish for high magnetic fields. Thus, any explanation in terms of AAS would imply
that a high magnetic field induces the squeezing of the wires leading to an array of one dimensionnal leads. A
second explanation for the frequency doubling could be the electron-electron interactions which could induce
charge doubling. Up to now the origin of this $h/2e$ peak at high magnetic fields remains unclear.

In the present paper we report on new experiments on the cage effect. These experiments in an other type of sample
confirm both the existence of the cage effects and the systematic presence of the frequency doubling for high
magnetic fields.  We have realised \T3 lattices on a GaAlAs/GaAs  heterostructure on which  electrostatics gates
have been deposited in order to control the number of channels per wire. The fabrication process and the
experimental results concerning these devices are introduced  in section $2$. Finally,  in section $3$ we present
theoretical calculations concerning the influence of the disorder on the cage effect.

\section{Cage effect versus the number of channels}

\begin{figure}
\centerline{\epsfxsize=80mm
\epsffile{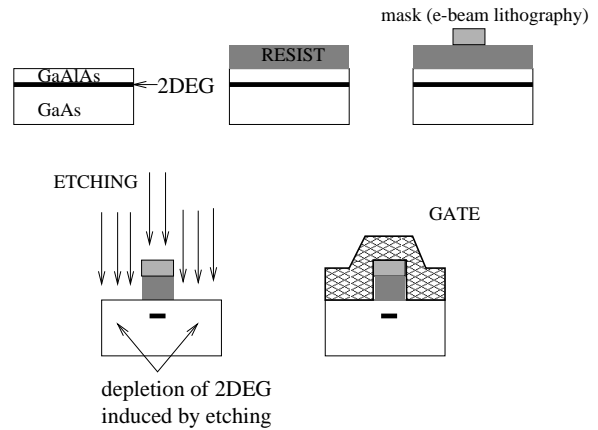}}
\vspace{3mm}
\caption{Step by step process for the realisation of the gated samples.}
\label{fab}
\end{figure}

This work is carried out on  high-mobility 2DEG heterojunction  material with an initial mobility of $3 \times
10^{6}$ cm$^{2}$ V$^{-1}$ s$^{-1}$ and a carrier density of $3.7 \times 10^{11}$ cm$^{-2}$ which give a Fermi
wavelength $\lambda_{F}=41$ nm and an elastic mean free path $l_{e}=12$ $\mu$m. Figure (\ref{fab}) shows the
experimental procedure to realize the electrostatic gate.
%
%
%%%%%%%%%%%%%%%%%%%%%%%%%%%%%%%%%
%%%%%%%%%%%%%%%%%%%%%%%%%%%%%%%%%
%
%
%
%
%%%%%%%%%%%%%%%%%%%%%%%%%%%%%%%%%
\begin{figure}
\centerline{\epsfxsize=80mm
\epsffile{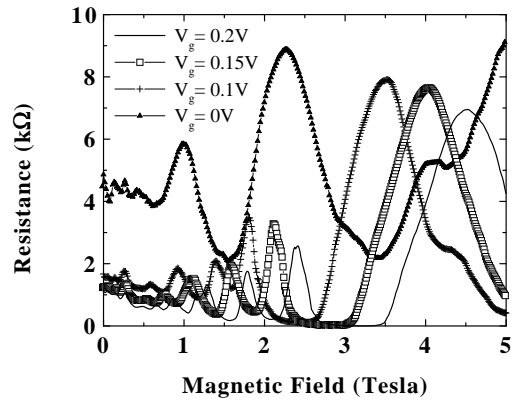}}
\caption{Magnetoresistance of a \T3 lattice versus the gate voltage showing
Shubnikov-de-Haas oscillations.}
\label{shub}
\end{figure}
%%%%%%%%%%%%%%%%%%%%%%%%%%%%%%%%%
%
%
A 200 nm thick insulator organic resist is first deposited on the top of the sample.  We use then electron beam
lithography and a lift-off technique to pattern the \T3 lattice with an aluminium mask (see figure
(\ref{lattice})). Reactive ion etching removes the resist from the  unprotected regions. Then, etching using Argon
ions transfers the pattern into the 2DEG. Due to the lateral depletion resulting  from the etching process, the
effective electrical width is considerably smaller than the wire width mask.   The metal of the gate is then
evaporated onto the whole  surface. As a result, the gate is in intimate contact with the regions where the  2DEG
has been destroyed whereas the thick resist layer considerably reduces the influence of the gate on the gas
beneath the wires. As a consequence, using this approach, we reduce the gate action on the electron density and
favour a pinching  of the wires making the lattice.
 In figure (\ref{shub}), we present the magnetoresistance of the \T3 network for several gate voltage values. The
measurements are performed at $50$ mK. From the Shubnikov-de-Haas oscillations we can extract the electron
density  versus the gate voltage.
%
%
%%%%%%%%%%%%%%%%%%%%%%%%%%%%%%%%%
\begin{figure}
\centerline{\epsfxsize=80mm
\epsffile{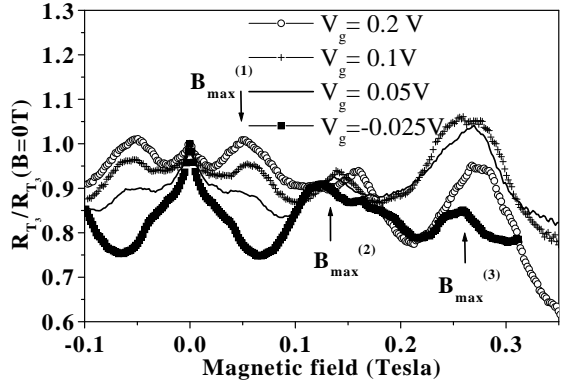}}
\vspace{4mm}
\caption{Same as figure (\ref{shub}) but in the low magnetic field range dominated 
by the classical boundary scattering effects.}
\label{class}
\end{figure}
%%%%%%%%%%%%%%%%%%%%%%%%%%%%%%%%%
%
%
%
%
%%%%%%%%%%%%%%%%%%%%%%%%%%%%%%%%%
\begin{figure}
\centerline{\epsfxsize=70mm
\epsffile{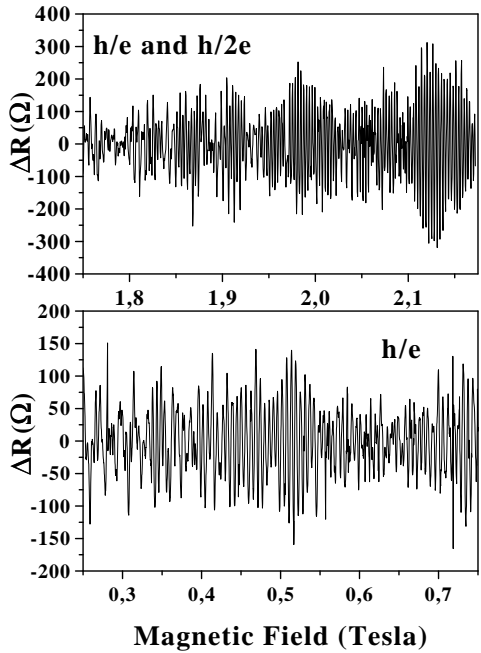}}
\caption{$h/e$ and $h/2e$ oscillations after substracting the background.}
\label{meb}
\end{figure}
%%%%%%%%%%%%%%%%%%%%%%%%%%%%%%%%%
%
%

In the low magnetic field regime we clearly observe large features  which depend on the gate voltage (indicated
by arrows in figure (\ref{class})). These features appear already at a temperature of 4K and are almost
temperature independent down to 50 mK, indicating a classical effect.  In 2DEG systems, the mean free path at low
temperature can be much larger than the width of the wire and the electrons are frequently scattered by the
boundaries.  Using a quasi-classical description, Ziman \cite{ziman} considers the scattering of electrons by a
rough  boundary. The conclusions agree with our intuitive expectations~: when the wavelength of the
incident electron is much smaller than the average boundary roughness, the electron is strongly
scattered in all directions. In the opposite  case, when the Fermi wavelength is much larger, the
incident electron undergoes specular reflections. These scatterings  at the boundaries \cite{thornton}
lead naturally to the concept of the boundary scattering length, $l_{b}$, which is the average distance
an electron can travel along a wire before a diffusive scattering event takes place. Then, the
effective mean free path $l_{eff}$ is such that
$\frac{1}{l_{eff}}=\frac{1}{l_{trans}}+\frac{1}{l_{b}}$ and, as long as $l_{trans}\gg l_{b}$, boundary scattering
dominates the momentum relaxation. The boundary scattering rate can be ``tuned'' using a magnetic field
perpendicular to the wire. At zero or low magnetic field, electrons with a large component of momentum parallel
to the boundaries contribute significantly to the  conductivity. However, as the magnetic field  increases, more
and more electrons are forced to interact with the edges and random scattering reduces the effective mean free
path $l_{eff}$. Thus, the resistivity increases and  subsequently saturates at some maximum value when the
cyclotron radius is approximatively twice the wire width \textrm{[9]}~:
\begin{equation} W =0.55 \times r_{c}(B_{max})= 0.55 \times \frac{\sqrt{2 \pi
\hbar^{2} n_{s}}}{e B_{max}}
\label{eq1}
\end{equation} Any further increase of the magnetic field will now lead to a drop of the resistivity since the
probability for an electron to be backscattered  by the edges is reduced. In figure (\ref{class}) we observe  $3$
maxima  corresponding to the $3$ magnetic field values denoted $B_{max}^{(1)}$, $B_{max}^{(2)}$ and 
$B_{max}^{(3)}$.  These maxima correspond to three different widths we can identify in our  samples~:
the width of the wires connecting the nodes, the diameter of the cavities formed at the sixfold
coordinated and the threefold coordinated lattice sites. Using equation (\ref{eq1}) we can deduce the
real dimension of the electron path. The channel number can be deduced from the comparison between the
Fermi wavelength and the wire width ($N=\frac{W^{(3)}}{\lambda_{F} /2}$).
These results are reported Table \ref{Table1}.

The identification of $W^{(3)}$ to the wire width gives reasonable numerical values. Nevertheless, the values
obtained for the cavities diameters are  overestimated, which is not surprising since this wire model is not
really adapted to the particular shape of the junctions.

The unexpected frequency doubling leads us to define two magnetic  field regimes~: a \textit{low field}
regime  (between $0.4$T and $1.2$T) where we observe only $h/e$ oscillations and a \textit{high field}
regime (from
$1.2$T to $2$T) where the $h/2e$ oscillations appear in addition to the $h/e$ ones.  We substract a polynomial
fit of the global magnetoresistance to the experimental points in order to extract the oscillations. This
resulting signal (see figure (\ref{meb})) is then Fourier transformed.

%
%
%%%%%%%%%%%%%%%%%%%%%%%%%%%%%%%%%
\begin{figure}
\vspace{5mm}
\centerline{\epsfxsize=80mm
\epsffile{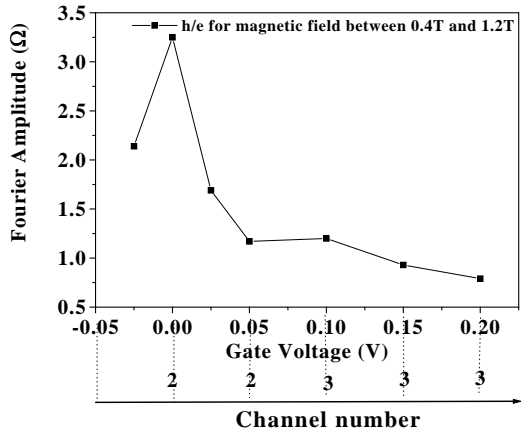}}
\vspace{4mm}
\caption{Variation of the $h/e$ peak amplitude of the Fourier transform versus the channel number for the \textbf{low magnetic field regime}.}
\label{low}
\end{figure}
%%%%%%%%%%%%%%%%%%%%%%%%%%%%%%%%%
%
%
%
%
%%%%%%%%%%%%%%%%%%%%%%%%%%%%%%%%%
\begin{figure}
\centerline{\epsfxsize=80mm
\epsffile{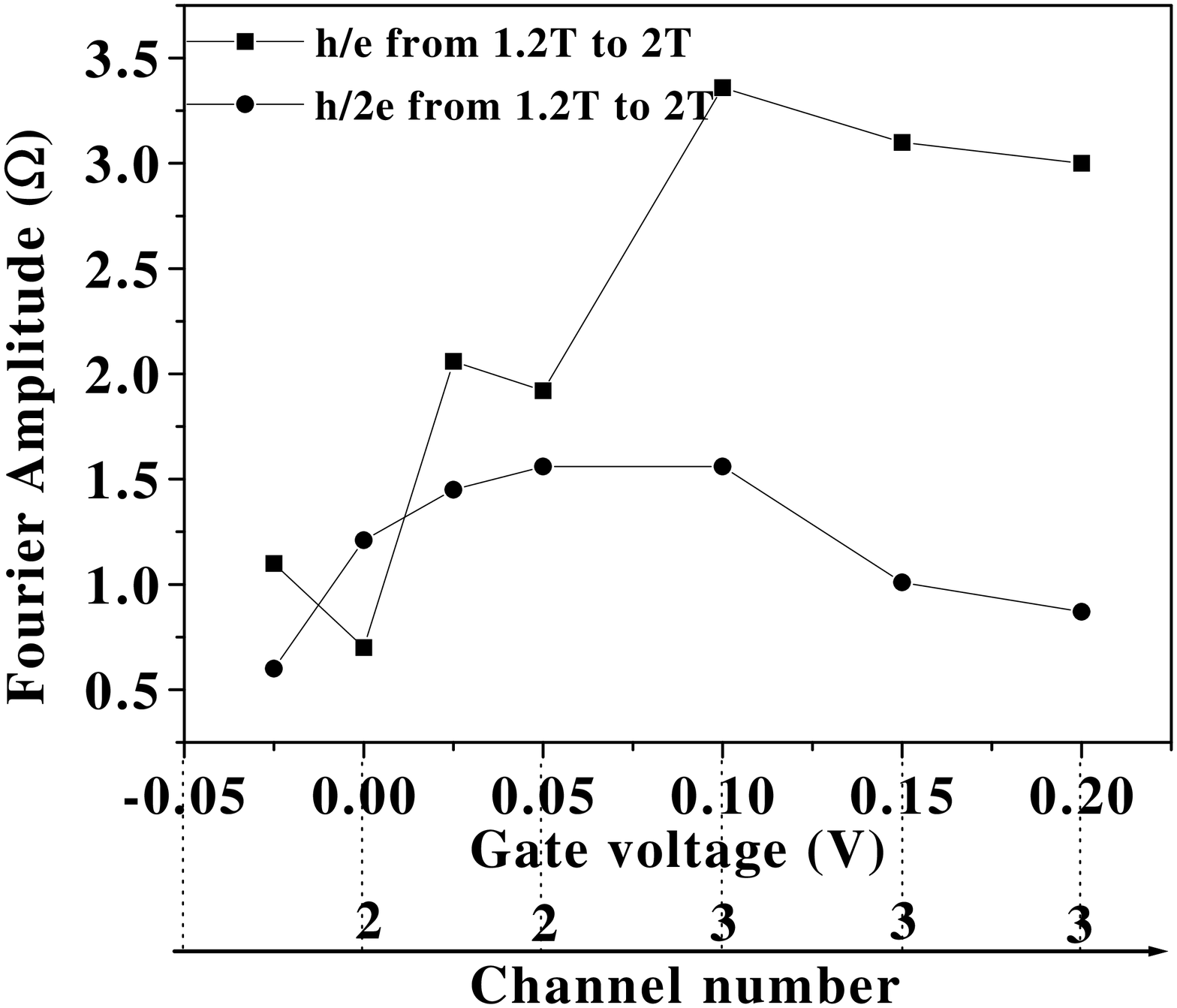}}
\vspace{4mm}
\caption{Variation of the $h/e$ and $h/2e$ peak amplitudes of the Fourier
transform  versus the  channel number for the \textbf{high magnetic field regime}.}
\label{high}
\end{figure}
%%%%%%%%%%%%%%%%%%%%%%%%%%%%%%%%%
%
%

Let us now discuss the gate voltage influence on the oscillations in  both regimes which is reported in figures
(\ref{low}) and (\ref{high}), where we have plotted the Fourier transform peak magnitude against the channel
number. In the low field regime the AB amplitude clearly increases with decreasing the electrical wire width.
This is in agreement with the work of M.~B\" uttiker
\textit{et al.} on a simple loop \cite{imry}. Essentially, their  calculation shows that if $N$ is the channel
number, the $h/e$ signal results from $N^{2}$ uncorrelated contributions whose stochastic averaging gives rise to
a decrease of the AB oscillation amplitude with $N$.  Note that in our experiments the amplitude of the Fourier
transform falls down at a critical gate voltage (about $0V$ corresponding to a channel number close to $2$). This
may be due to the wire pinch-off induced by the gate. As a result, the cutting of some wires supresses the cage
effect and consequently reduces the $h/e$ signal. In the high field regime we observe a different behavior
between the 
$h/e$ and the $h/2e$ periods. The $h/e$ amplitude decreases monotonously with the reduction of the channel
number. On the other hand, the $h/2e$ oscillations seem to have a constant magnitude versus the gate voltage.
This fact suggests a different mechanism to explain the two periodicities in this high field regime. More
surprisingly, the $h/e$ signal itself behaves in an opposite manner in the two regimes~: whereas, as
discussed above, the AB amplitude at low field has the expected channel number dependence, in the high
field regime the
$h/e$ signal increases with $N$. This leads us to raise the hypothesis that we are  dealing with different
physical phenomena in the two magnetic field regimes. Up to now the origin of the $h/2e$ peak at high magnetic
fields and, more, the field strength effect on the $h/e$ peak amplitude, remain unclear.

\section{The effect of disorder} In this section, we discuss the influence of an elastic disorder on the
periodicity of the AB oscillations in the \T3 lattice. In  order to simplify this analysis, we consider that
there exists only one conduction channel. Of course this hypothesis is somehow unrealistic but it allows us to
get a partial answer to this problem. The technical details can be found in ref.~\textrm{[11]}.

\begin{figure}
\centerline{\epsfxsize=80mm
\epsffile{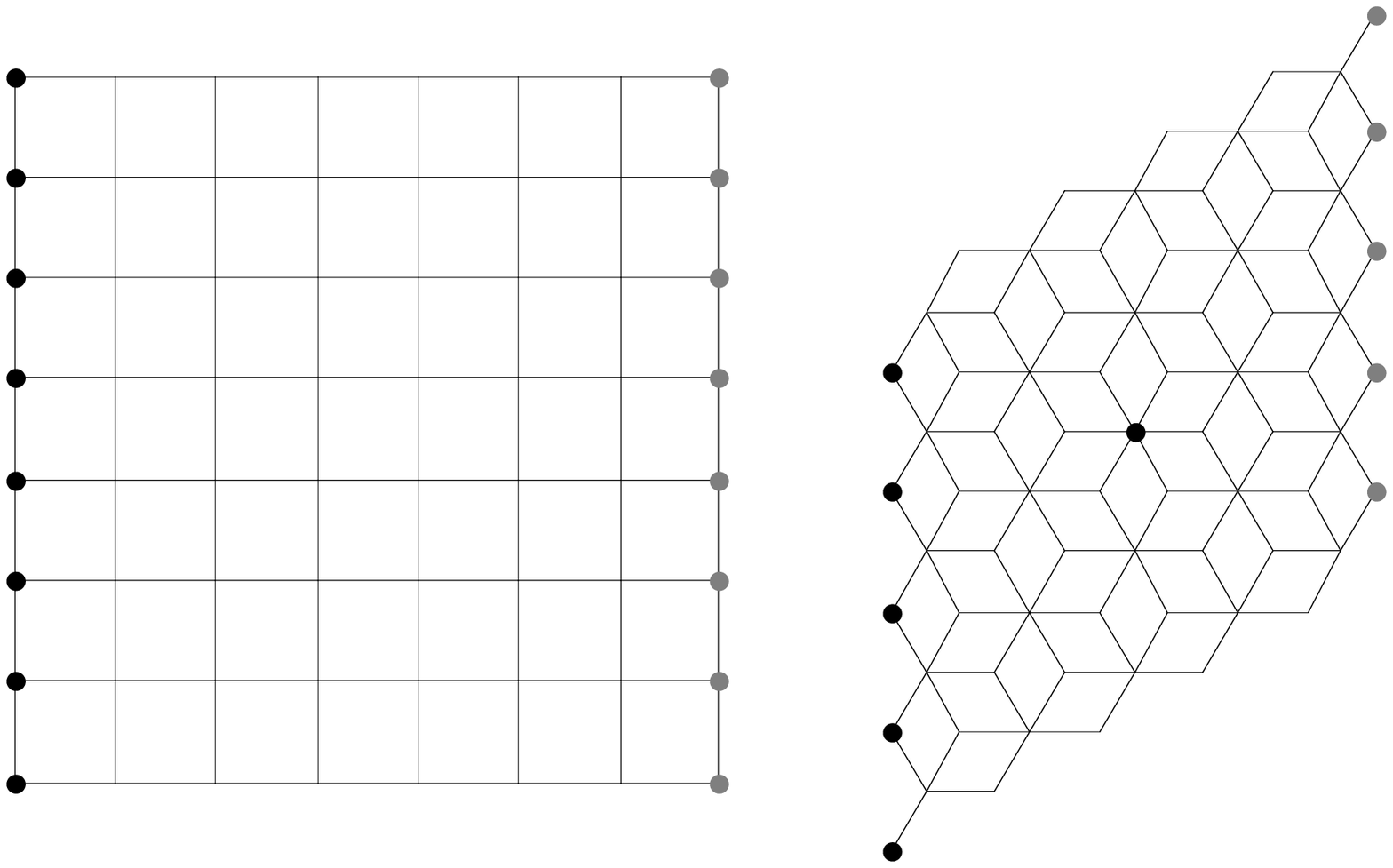}}
\vspace{4mm}
\caption{A piece of the square lattice (left) and of the \T3 network (right).
Black (respectively grey) dots represent the connections to the input  (respectively output) channels. 
The central black dot is the input  channel chosen for the  bulk injection.}
\label{syst}
\end{figure}

 From a theoretical point of view, there are several ways to introduce  disorder in a system. For instance, one
can consider a random scattering matrix on each bond of the network or at each node. We can also put some
additionnal  dead-end bonds of random lengths to mimick the dephasing due to impurities. Here, we adopt another
point of view by changing the length of each quantum path. One way to do this practically, is to make a random
modulation $\Delta l$ of each bond of length $l$ without modifying the phase factor due to the magnetic field.
Indeed, since each bond is  considered as strictly one-dimensional, the circulation of the vector potential along
one link is not affected by reflections of the wavepacket on this link. Thus, in the Landauer-like transmission
formalism that we have used, we only include these length fluctuations in the quantity $kl$ which is the main
parameter of our study \cite{vidal2}. In the following, we focus on the transmission properties of a piece of the
\T3 lattice and we compare it to a piece of the square lattice (see figure (\ref{syst})).
%
%
%%%%%%%%%%%%%%%%%%%%%%%%%%%%%%%%%
%%%%%%%%%%%%%%%%%%%%%%%%%%%%%%%%%
%
%

\begin{figure}
\centerline{\epsfxsize=80mm
\epsffile{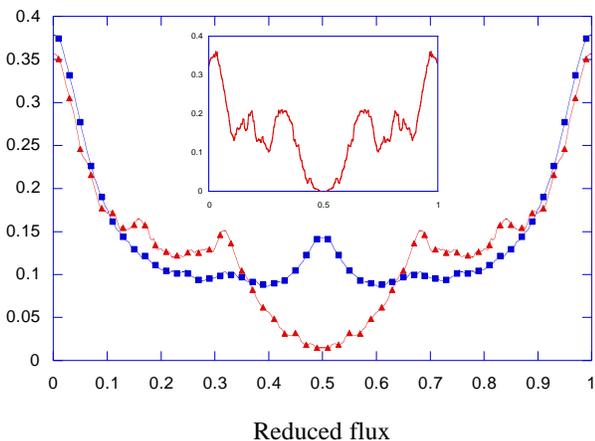}}
\vspace{4mm}
\caption{Averaged transmission coefficient  as a
function of the reduced flux for the square lattice (square)  and for 
the \T3 network (triangle). Inset~: averaged transmission coefficient for the
\T3 network with a single input channel in the bulk of the network.}
\label{trans}
\end{figure}

In the pure case ($\Delta l=0$) the transmission coefficient $T$  depends on the wave vector $k$ of the incoming
wave function and on the reduced flux $f$. We have displayed in figure (\ref{trans}), the averaged transmission
coefficient for $kl \in [0,2\pi]$. One clearly observes some peaks for rationnal $f$ (extended Bloch-like
eigenstates) reminiscent of the butterfly-like structure of the  energy spectrum. Due to the existence of the AB
cages, the transmission coefficient is minimum at 
$f=1/2$ for the \T3 network but it does not strictly vanish. This is due to the  existence of dispersive edges
states \cite{vidal3} that are able to carry current even for $f=1/2$.  Nevertheless, when one injects the current
in the bulk of the sample, we obtain $T=0$ for $f=1/2$ as expected.  Note that for both systems, the signal is
anharmonic but is periodic with period $\phi_0$ leading to a $h/e$ periodicity for the magneto-resistance. 

%
%
%%%%%%%%%%%%%%%%%%%%%%%%%%%%%%%%%
%%%%%%%%%%%%%%%%%%%%%%%%%%%%%%%%%
%
%
%
%
%%%%%%%%%%%%%%%%%%%%%%%%%%%%%%%%%
\begin{figure}
\centerline{\epsfxsize=80mm
\epsffile{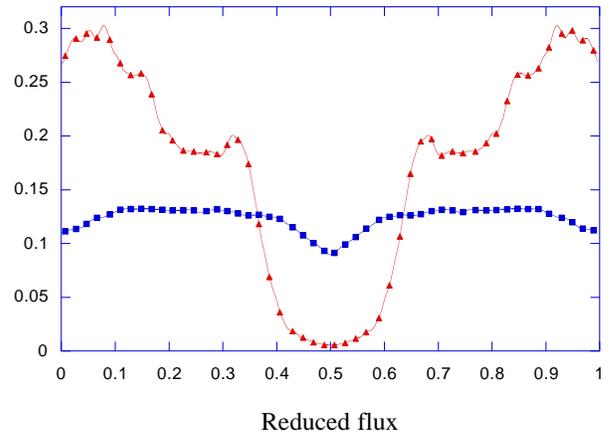}}
\vspace{4mm}
\caption{Transmission coefficient averaged over 50 configurations of 
disorder for $k l=\pi/3$
and $k\Delta l = 1.47$ as a function of the reduced flux.}
\label{desordre}
\end{figure}
%%%%%%%%%%%%%%%%%%%%%%%%%%%%%%%%%
%
%

The main question is to know how this periodicity is modified when we introduce the disorder. It is clear that if
the  disorder is very strong ($k\Delta l \simeq \pi$), there is no phase coherent transmission and we expect a
vanishing transmission after averaging over disorder. Consequently,  there might be interesting phenomenon for
weak enough disorder. We emphasize that the notion of weak and strong disorder is related  to the product $k
\Delta l$ and thus depends on $k$. Note that this way to model a static disorder is not expected to capture all
the features of experimental data. For instance, it yields an effective disorder which increases with the Fermi
energy, but this effect could well be completely compensated by the energy dependence of the impurity scattering
cross section in real systems. We have shown in figure (\ref{desordre}) the effect of disorder on the 
transmission coefficient as a function of the magnetic flux. It is clearly seen that for the square network, the
periodicity is no  longer $\phi_0$ but $\phi_0/2$. This new periodicity is simply due to phase coherent pairs of 
time-reversed trajectories according to the weak-localization picture. The most  striking feature is that for the
\T3 lattice, the transmission coefficient remains  $\phi_0$-periodic with a large amplitude. This strongly
suggests that the cage effect (which locks  the phase of the oscillations) survives for this strength of
disorder. Physically, it means that a strong dephasing is required to destroyd the completely destructive quantum
interferences responsible of the AB cages. In addition, note that, for this energy,  the signal of the \T3
lattice transmission coefficient  is about 10 times larger than the  one of the square network. In ref.
\cite{vidal3} we report on other energy of the incoming wavepacket where we find always the same type of feature.

\section{Conclusions}

Localisation phenomenon induced by the magnetic field on the \T3 topology has been confirmed in the GaAlAs/GaAs
system. The frequency doubling in the high magnetic field regim is systematically observed and displays a
different behavior versus the channel number compared to the fundamental period. The theoretical predictions of
the effect of disorder motivate the investigation of such networks in metallic systems.

%%%%%%%%%%%%%%%%%%%%%%%%%%%%%%%%%%%%%%%%%%%%%%%%%%%%%%%%%%%%%%%%%%%

\begin{widetext}
\begin{center}
Table I~: 
\end{center}

\begin{center}
\begin{tabular}{lllllllll} 
Gate  & $n_{s}$ & $B_{max}^{(1)}$ & $W^{(1)} $ &$ B_{max}^{(2)}$ & $W^{(2)}$& $B_{max}^{(3)}$ & $W^{(3)} $ & N  \\
voltage & $10^{11} cm^{-2}$ & (Tesla) & ($\mu m$) & (Tesla) & ($\mu m$) & (Tesla) & ($\mu m$)  \\
\hline
\hline
$V_{g}=0.2 V$&1.48 & 0.052                   & 0.67                          &
0.159 & 0.219 & 0.274         &  0.127&3               \\
$V_{g}=0.15 V$&1.32&0.052                   &0.63                           &
0.154 & 0.219 &  0.268    &      0.123&3          \\
$V_{g}=0.1 V$&1.17& 0.056                   &  0.56 
&0.140
& 0.221  &  0.262   &  0.118&3            \\
$V_{g}=0.05 V       $&0.9&        hard to           & locate       &0.137&  0.199      & 0.270 & 0.101&2  \\
$V_{g}=0V $&0.83     &  hard to   & locate  &  0.122 & 0.213 & 0.256 & 0.101&2  \\
\hline
\hline
\label{Table1}
\end{tabular}
\end{center}
\end{widetext}


\begin{thebibliography}{10}

\bibitem{hof} D.~R.~Hofstadter, \PRB  \textbf{14}, 2239 (1976).

\bibitem{vidal}J.~Vidal, R.~Mosseri, and B.~Dou\c cot,
\PRL \textbf{81}, 5888 (1998).


\bibitem{pannetier}C.~C.~Abilio, P.~Butaud, Th.~Fournier, and B.~Pannetier,
J.~Vidal, S.~Tedesco and B.~Dalzotto,
\PRL \textbf{83}, 5102 (1999).

\bibitem{pannetier2}E.~Serret \textit{et al.}, in preparation.

\bibitem{naud}C.~Naud, G.~Faini, and D.~Mailly, 
\PRL \textbf{86}, 5104 (2001).

\bibitem{wasburn}F.~P.~Milliken, S.~Wasburn, C.~P.~Umbach, R.~B.~Laibowitz, and R.~A.~Webb, 
\PRB  \textbf{36}, 4465 (1987).

\bibitem{pepper}T.~J.~Thornton, M.~Pepper, H.~Ahmed, D.~Andrews, and G.~J.~Davies,
\PRL  \textbf{56}, 1198 (1986).


\bibitem{ziman}J.~M.~Ziman,
\textit{Electrons and phonons} Chap. 11, Oxford Univ. Press UK (1960).

 \bibitem{thornton}T.~J.~Thornton, M.~L.~Roukes, A.~Scherer, and B.~P.~E. Van der Gaag,
{\it Coherence in Mesoscopic Systems}, edited by B.~Kramer, Plenum Press (New York and London), 
series B~: Physics Vol. \textbf{254}, 153 (1990).

\bibitem{imry}M.~B\" uttiker, Y.~Imry, R.~Landauer, and S.~Spinhas, 
\PRB \textbf{31}, 6207 (1985).


\bibitem{vidal2}J.~Vidal, G.~Montambaux, and B.~Dou\c cot, 
\PRB \textbf{62}, R16294 (2000).

\bibitem{vidal3} J.~Vidal, P.~Butaud, B.~Dou\c cot, and  R.~Mosseri,
to be published in Phys. Rev. B.

\end{thebibliography}
\end{document}